\documentclass[11pt, fleqn]{article}
\usepackage{amsmath}
\usepackage{booktabs}
\usepackage{amssymb}
\usepackage{latexsym}

\usepackage[dvips,dvipdfmx]{graphicx,color}

\begin{document}

\hfill MISC-2016-07

\begin{center}
{\Large\bf Muon-Electron Conversion in a Family Gauge Boson Model}

\vspace{4mm}
{\bf Yoshio Koide$^a$ and Masato Yamanaka$^b$  }

${}^a$ {\it Department of Physics, Osaka University, 
Toyonaka, Osaka 560-0043, Japan} \\
{\it E-mail address: koide@kuno-g.phys.sci.osaka-u.ac.jp}

${}^b$ {\it Maskawa Institute,
 Kyoto sangyo University, 
Kyoto 603-8555, Japan } \\
{\it E-mail address: masato.yamanaka@cc.kyoto-su.ac.jp}

\end{center}

\vspace{3mm}

\begin{abstract}
We study the $\mu$-$e$ conversion in muonic atoms via an 
exchange of family gauge boson (FGB) $A_{2}^{\ 1}$ in a 
$U(3)$ FGB model. Within the class of FGB model, we consider 
three types of family-number assignments for quarks. 
We evaluate the $\mu$-$e$ conversion rate for various target 
nuclei, and find that next generation $\mu$-$e$ conversion 
search experiments can cover entire energy scale of the model 
for all of types of the quark family-number assignments. 
We show that the conversion rate in the model is so sensitive 
to up- and down-quark mixing matrices, $U^{u}$ and $U^{d}$, 
where the CKM matrix is given by $V_\text{CKM} = U^{u\dagger} 
U^d$. 
Precise measurements of conversion rates for various target nuclei 
can identify not only the types of quark family-number assignments, 
but also each quark mixing matrix individually. 
\end{abstract}

\

PCAC numbers:  
  11.30.Hv, 
  12.60.-i, 
  14.60.Ef, 
  14.70.Pw, 

\newpage
\section{Introduction}

The idea of family gauge bosons (FGBs) 
$A_i^{\ j}$ ($i,j=1,2,3$) seems to be the most 
natural extension of the standard model (SM). 
In the SM of quarks and leptons, a degree of freedom which is not 
yet accepted as a gauge symmetry is only that of the families 
(generations). So far, because of the severe constraint from the 
observed $P^{0}$-$\bar P^{0}$ mixing ($P=K, D, B, B_{s}$)
it has been considered that a scale of the 
FGBs is very large so that we cannot observe those at terrestrial 
experiments.

Against such conventional models, a FGB model on a low energy 
scale has been proposed by Sumino~\cite{Sumino_PLB09, 
Sumino_JHEP09}. The model has the following characteristics 
(details are given in Sec.~2): 
(i) Family symmetry $U(3)$ is broken at a low energy scale of 
$\mathcal{O} (10^3)\,\text{TeV}$. 
(ii) The FGB mass matrix is diagonal in the flavor basis in which 
the charged lepton mass matrix is diagonal, so that lepton 
family-number violation does not occur. 
(iii) FGB masses and gauge coupling $g_F$ are not free 
parameters. FGB masses and $g_F$ are related to the charged 
lepton masses and to the electroweak gauge coupling, 
respectively.  Hence the predictions in the model are less ambiguous.

There is a variety of types of FGB spectrum and quark 
family-number assignments. We focus on three models 
compatible with observed $P^{0}$-$\bar P^{0}$ mixing.  
In Model A, FGB masses have an inverted hierarchy, i.e., 
lightest and heaviest FGB are $A_{3}^{\ 3}$ and $A_{1}^{\ 1}$, 
respectively~\cite{K-Y_PLB12}. 
In Model B, the quark family-number is assigned as twisted, 
e.g., $(d_1,d_2, d_3)=(b, s, d)$ (Model B$_1$), and 
$(d_1,d_2, d_3)=(b, d, s)$ (Model B$_2$) for $(e_1,e_2, e_3) 
\equiv (e^-, \mu^-, \tau^-)$~\cite{Koide_PLB14}.

Our interest is in how to confirm the FGB model at terrestrial 
experiments. We have already pointed out a possibility that we 
observe the lightest FGB $A_1^{\ 1}$ in Model B at the 
LHC~\cite{K-Y-Y_PLB15}. 
There is however still a possibility that the FGB is too heavier 
to observe at the LHC. Now it is worth investigating how to 
check such too heavy FGBs.

In this paper, we focus on $\mu$-$e$ conversion in 
muonic atoms. The FGB $A_{2}^{\ 1}$ possesses a muon- and 
electron-number violating interaction, and gives rise to the 
$\mu$-$e$ conversion, but not other muon-number violating 
decays. 
New experiments to search for the $\mu$-$e$ conversion will 
launch soon, e.g., DeeMe, COMET, Mu2e, and PRISM 
experiment, whose single event sensitivities are $\text{B} 
({\rm Si}) \sim 5 \times 10^{-14}$ (DeeMe)~\cite{DeeMe}, 
$\text{B}({\rm Al}) \sim 3 \times 10^{-17}$ (COMET and 
Mu2e)~\cite{COMET, Mu2e}, and $\text{B}({\rm Al}) \sim 
7 \times 10^{-19}$ (PRISM)~\cite{COMET}. 
Here $\text{B}(N)$ denotes branching ratio of the 
$\mu$-$e$ conversion with a target nucleus $N$. 
We evaluate the $\mu$-$e$ conversion rate, and show 
that these experiments scan entire parameter space of the model. 
Once $\mu$-$e$ conversion events are discovered, we need 
to find out the $A_{2}^{\ 1}$ contribution in the events 
without relying on other muon-number violating observables. 
And, with only the $\mu$-$e$ conversion signals, we need 
to discriminate the FGB model from other models in which the 
$\mu$-$e$ conversion is dominant muon-number violating 
process~\cite{Dinh:2012bp, Alonso:2012ji, Toma:2013zsa, 
Abada:2014kba, Sato:2014ita}. 
We discuss the discrimination through the measurement 
of the branching ratios for various nuclei.

The precise measurement of the branching ratios plays an 
important role. The branching ratios in the model are sensitive  
to the quark mixing matrices $U^{u}$ and $U^{d}$, where 
Cabibbo-Kobayashi-Maskawa (CKM) matrix is given by 
$V_\text{CKM} = U^{u \dagger} U^{d}$~\cite{CKM}. 
We have chance to individually determine $U^{u}$ and $U^{d}$ 
through the measurements of the $\mu$-$e$ conversion. 
We will discuss the feasibility of it.

This work is organized as follows. First we briefly review the 
FGB model. We illustrate three types of quark family-number 
assignments. 
Then we introduce four types of quark mixing to describe the 
interaction between FGBs and quarks. 
Next, in Sec.~3, we formulate the $\mu$-$e$ conversion rate 
in the FGB model. 
In Sec.~4, we give numerical results, and show that the FGB 
model can be confirmed or ruled out at $\mu$-$e$ conversion 
search experiments in near future. We discuss feasibility for 
discriminations among three types of quark family-number 
assignments and four types of quark mixing matrices. 
Sec.~5 is devoted to summarize this work.

\section{Family gauge boson Model}

Let us give a brief review of a $U(3)$ family gauge boson (FGB) 
model proposed by Sumino~\cite{Sumino_PLB09}. 
Sumino has noticed a problem in a charged lepton mass 
relation~\cite{K-mass}, 
$$ 
K \equiv \frac{m_e + m_\mu +m_\tau}{\left(\sqrt{m_e}+ \sqrt{m_\tau} 
+\sqrt{m_\tau}\right)^2} = \frac{2}{3} . 
\eqno(2.1)
$$
The relation is satisfied by the pole masses, $K^{pole}=(2/3) 
\times(0.999989 \pm 0.000014)$, but not so well satisfied by the
running masses, $K(\mu = m_Z)=(2/3) \times (1.00189 \pm 
0.00002)$. 
The running masses $m_{e_i}(\mu)$ are given by~\cite{Arason92}
$$
m_{e_i}(\mu) = m_{e_i} \left[ 1-\frac{\alpha_{em}(\mu)}{\pi} 
\Bigl( 1 +\frac{3}{4} \log \frac{\mu^2}{m_{e_i}^2(\mu)} \Bigr) \right] .
\eqno(2.2)
$$
In the absence of family-number dependent factor 
$\log(m_{e_i}^2)$, the running masses $m_{e_i}(\mu)$ also 
satisfy the relation (2.1). In order to understand this puzzle, 
Sumino has proposed a $U(3)$ FGB model so that a factor 
$\log(m_{e_i}^2)$ from the QED correction is canceled by the 
FGB loop contribution $\log (M_{ii}^2)$~\cite{Sumino_PLB09}. 
Here, the masses of FGBs $A_i^{\ j}$, $M_{ij}$, are given by
$$
M_{ij}^2 = k (m_{e_i}^n + m_{e_j}^n),
\eqno(2.3)
$$
where $k$ is a constant with dimension of (mass)$^{2-n}$. 
The cancellation mechanism holds for any $n$, 
because $\log M_{ii}^n =n \log M_{ii}$. The original model 
has studied the $n=1$ case~\cite{Sumino_PLB09}. 
The cancellation requires the following relation between the 
family gauge coupling $g_F$ and QED coupling $e$,
$$
\left( \frac{g_F}{\sqrt2} \right)^2 =\frac{2}{n} e^2 =
\frac{4}{n} \left( \frac{g_w}{\sqrt2} \right)^2
\sin^2 \theta_w .
\eqno(2.4)
$$
Here $\theta_w$ is the Weinberg angle. 
Note that the cancellation mechanism holds only at the one loop level. 
Sumino has speculated the scale of $U(3)$ family symmetry is an 
order of $10^3$\,TeV \cite{Sumino_PLB09, Sumino_JHEP09}.

In the FGB model, the family symmetry is broken by a scalar 
$\Phi$ with $({\bf 3}, {\bf 3})$ of $U(3) \times O(3)$. 
The family-numbers of quarks and leptons, which are triplets of 
$U(3)$, are changed only by exchanging $\Phi\bar{\Phi}$, not 
the single $\Phi$. Thus, the FGB contribution to 
pseudo-scalar 
meson oscillations is highly suppressed. 
The FGB mass matrix is diagonal in the flavor basis in which 
the charged lepton mass matrix is diagonal, because those masses 
are generated by the common scalar $\Phi$. 
Therefore, family-number violation does not occur in the charged 
lepton sector.

In the original model, charged leptons $(e_{Li}, e_{Ri})$ 
are assigned to $({\bf 3}, {\bf 3}^*)$ of $U(3)$ family symmetry, 
which makes the sign of FGB loop correction to be opposite to 
the QED correction for the cancellation. So the original model is 
not anomaly free. 
In order to avoid this anomaly problem, Yamashita and one of the 
authors (YK) have proposed an extended FGB model~\cite{K-Y_PLB12}: 
two scalars $\Psi$ and $\Phi$ are introduced, which are 
$({\bf 3}, {\bf 3}^{*})$ of $U(3) \times U(3)'$. 
Charged lepton masses are generated via the VEV of $\Phi$ only. 
FGB masses are achieved via the VEVs of $\Phi$ and $\Psi$. 
Relations of these VEVs are $\langle \Psi \rangle \propto \langle 
\Phi \rangle^{-1}$ and $\langle \Psi \rangle \gg \langle \Phi 
\rangle$. These relations lead the FGB spectrum (2.3) with 
negative $n$, in contrast to the original FGB model in which a VEV 
of single scalar field generates both of masses of charged leptons and 
FGBs. 
We can therefore realize the cancellation with a normal 
assignment $(e_{Li}, e_{Ri}) =({\bf 3}, {\bf 3})$ of $U(3)$ 
family symmetry, because of  $\log M_{ii}^n = n \log M_{ii} <0$ 
with the negative $n$.

In this paper, we call the extended FGB model Model A, and 
call the original model Model B.
The characteristics of these models are summarized in Table 1. 
In order to relax the severe constraints from the observed 
$P^0$-$\bar{P}^0$ mixings, we consider that the lightest FGB 
interacts with only the third generation quarks. We define the 
family-number as $(e_1, e_2, e_3)=(e^-, \mu^-, \tau^-)$.
In Table 1, we list ``optimistic" lower limit on $M_{12}$ 
which is not conflict with all of observed $P^0$-$\bar{P}^0$ 
mixings~\cite{Koide_PLB14}.

\begin{table}[t]
\caption{Three extended FGB models. 
$q^{0}$ stands for eigenstates of the $U(3)$ family gauge 
symmetry. Note that this lower bound on $M_{12}$ is 
derived from $P^{0}$-$\bar{P}^{0}$ mixing 
measurements~\cite{Koide_PLB14}, not from $\mu$-$e$ 
conversion search experiments. }
{\footnotesize
\vspace{2mm}
\begin{tabular}{llll}\hline
& Model A  
& Model B$_1$ 
& Model B$_2$ 
\\ \hline
Symmetries   
& $U(3) \times U(3)'$ 
& $U(3) \times U(3)'$
& $U(3) \times U(3)'$
\\
lepton currents   
& $\bar{\ell}^i\gamma_\mu \ell_j$ 
& $\bar{\ell}^i_L\gamma_\mu \ell_{jL} - 
\bar{\ell}_{jR}\gamma_\mu \ell^i_R$
& $\bar{\ell}^i_L\gamma_\mu \ell_{jL} - 
\bar{\ell}_{jR}\gamma_\mu \ell^i_R$ 
\\ 
quark currents  
& $\bar{q}^{0i} \gamma_\mu q^0_{j}$ 
& $\bar{q}^{0i} \gamma_\mu q^0_{j}$
& $\bar{q}^{0i} \gamma_\mu q^0_{j}$
\\
$g_F/\sqrt{2}$  
& $0.491/\sqrt{n}$   
& $0.428/\sqrt{n}$
& $0.428/\sqrt{n}$
\\
$(e_1, e_2, e_3)$ 
& $(e^-, \mu^-, \tau^-)$ 
& $(e^-, \mu^-, \tau^-)$
& $(e^-, \mu^-, \tau^-)$
\\
$(d_1, d_2, d_3)$ 
& $(d^0, s^0, b^0)$ 
& $(b^0, s^0, d^0)$ 
& $(b^0, d^0, s^0)$ 
\\
$M_{11} : M_{22} : M_{33} $
& $(1/m_e)^{n/2} \hspace{-1.5pt}:\hspace{-1.5pt} 
(1/m_\mu)^{n/2} \hspace{-1.5pt}:\hspace{-1.5pt} 
(1/m_\tau)^{n/2}$ 
& $m_e^{n/2} \hspace{-1.5pt}:\hspace{-1.5pt} 
m_\mu^{n/2} \hspace{-1.5pt}:\hspace{-1.5pt} 
m_\tau^{n/2}$ 
& $m_e^{n/2} \hspace{-1.5pt}:\hspace{-1.5pt} 
m_\mu^{n/2} \hspace{-1.5pt}:\hspace{-1.5pt} 
m_\tau^{n/2}$ 
\\
lower bound of $M_{12}$  ($n=1$)   
& $1.76 \times 10^3\,\text{[TeV]}$ 
& $98.4\,\text{[TeV]}$    
& $98.0\,\text{[TeV]}$ 
\\
lower bound of $M_{12}$ ($n=2$)    
& $1.80 \times 10^4\,\text{[TeV]}$ 
& $78.2 \,\text{[TeV]}$    
& $77.9\,\text{[TeV]}$ 
\\ \hline
\end{tabular} 
}
\end{table}

\subsection{Model A}

According to the extended FGB model, Model A is characterized by 
the following inverted mass hierarchy of FGB mass~\cite{K-Y_PLB12}, 
$$
M_{ij}^2 \propto \frac{1}{m_{ei}^{n}} + \frac{1}{m_{ej}^{n}} , 
\eqno(2.5)
$$
($n$ is a positive integer). Interaction Lagrangian of quarks and 
leptons with the FGBs is given by 
$$
{\cal L} = \frac{g_F}{\sqrt2}\left\{ \sum _{\ell=e, \nu} 
 (\bar{\ell}^i \gamma_\mu \ell_j) +\sum_{q=u, d} 
 U^{q *}_{ik} U^{q}_{jl} 
(\bar{q}^k \gamma_\mu q_{l}) \right\} (A^\mu)_i^{\ j}. 
\eqno(2.6)
$$
Here $q^0_i = U^{q}_{ij} q_j$ is  an interaction eigenstate 
of the $U(3)$ symmetry, where $q_j$ and $U^{q}_{ij}$ represent 
mass eigenstate and quark mixing matrix, respectively. 
The interactions are a type of pure vector, so that the model is 
anomaly free. The gauge coupling $g_F$ in Model A is given as 
\cite{K-Y_PLB12}  
$$
 \frac{g_F}{\sqrt2}  = 
\left[ \frac{3 \zeta}{2n} 4 \pi \alpha_{em}(m_\mu) \right]^{1/2} 
=  \frac{1}{\sqrt{n}} 0.491 ,
\eqno(2.7)
$$
where $\alpha_{em}(m_\mu)=1/137$, and  $\zeta = 
1.752$ is a fine tuning factor which is obtained from phenomenological 
study.

\subsection{Model B}

Model B is characterized by the following relation of FGB mass,
$$
M_{ij}^2 \propto m_{ei}^{n} + m_{ej}^{n} ,
\eqno(2.8)
$$
($n$ is a positive integer).
Interaction Lagrangian of quarks and leptons with the FGBs 
is given by 
$$
\begin{array}{ll}
\displaystyle{
{\cal L} = \frac{g_F}{\sqrt2}
\biggl\{ 
\sum _{\ell=e, \nu} 
\left[ 
(\bar{\ell}_L^i \gamma_\mu \ell_{Lj} ) 
-(\bar{\ell}_{Rj} \gamma_\mu \ell_R^i) 
\right] 
}
\displaystyle{ 
+\sum_{q=u, d} 
 U^{q*}_{ik} U^{q}_{jl} 
(\bar{q}^k \gamma_\mu q_{l}) 
\biggr\} 
(A^\mu)_i^{\ j}
}. 
\end{array}
\eqno(2.9)
$$
Here, note that the leptonic currents have an unfamiliar form, 
$(V-A)^i_{\ j} -(V+A)_j^{\ i}$, because fermions $(\psi_L, 
\psi_R)$ are assigned to $({\bf 3}, {\bf 3}^*)$ of $U(3)$.
Since this assignment in the quark sector leads unwelcome 
large $K^0$-$\bar{K}^0$ mixing, we use pure vector current 
form as far as quark currents are concerned.  
The gauge coupling $g_F$ is given by
$$
 \frac{g_F}{\sqrt2}  = 
\left[ \frac{2}{n} 4 \pi \alpha_{em}(m_\mu) \right]^{1/2} 
= \frac{1}{\sqrt{n}}\, 0.428 .
\eqno(2.10)
$$
In order to avoid the severe constraints from the observed 
$P^0$-$\bar{P}^0$ mixing, the lightest FGB $A_1^{\ 1}$ couples 
only to the third generation quarks, so that we have the following 
two scenarios for the family-number assignment\,\cite{Koide_PLB14}: 
$$
\begin{array}{ll}
(d_1, d_2, d_3) = (b^0, s^0, d^0)  & {\rm in\ Model\ B}_1 , \\
(d_1, d_2, d_3) = (b^0, d^0, s^0)  & {\rm in\ Model\ B}_2 .
\end{array}
\eqno(2.11)
$$

\subsection{Typical cases of quark mixing}

In the FGB model, $\mu$-$e$ conversion branching ratio 
$\text{B}(\mu^- N \rightarrow e^- N)$ is 
sensitive to the quark mixing matrices, $U^u$ and 
$U^{d}$. Each explicit form is not determined yet, 
though the combination is measured as $V_\text{CKM} = 
(U^{u})^\dagger U^{d}$. 
We calculate $\text{B}(\mu^- N \to 
e^- N)$ by using some typical mixing matrices from the 
practical point of view.

The family numbers do not always correspond to the generation 
numbers in Model B. In order to avoid confusing, hereafter, 
we denote $U^u$, $U^d$ and $V_{CKM}$ in the generation basis 
and, e.g., we denote $(U^d)_{12}$ as $(U^d)_{ds}$.

As the first case (Case I), we consider following mixing, 
$$ 
U^u  \simeq {\bf 1}, \ \ \ \  U^d \simeq V_\text{CKM}, 
\ \ \ \ \ \ \ \ ({\rm Case\ I}) . 
\eqno(2.15)
$$ 
Case I is the most likely case.  
Since we know $m_t/m_u \gg m_b/m_d$, we consider that 
the CKM mixing almost comes from down-quark mixing $U^d$. 
Besides, we know an empirical well-satisfied relation 
$V_{us} \simeq \sqrt{m_d/m_s}$ without 
$\sqrt{m_u/m_t}$~\cite{Vus}. 
In fact, Case I is practically well satisfied in most of mass 
matrix models. We adopt the standard expression for the 
explicit form of $V_\text{CKM}$, 
$$
V_\text{CKM} = \left(
\begin{array}{ccc}
c_{13} c_{12} & c_{13} s_{12} & s_{13} e^{-i\delta} \\
-c_{23} s_{12} -s_{23} c_{12} s_{13} e^{i\delta} &
c_{23} c_{12} -s_{23} s_{12} s_{13} e^{i\delta} &
s_{23} c_{13} \\
s_{23} s_{12} -c_{23} c_{12} s_{13} e^{i\delta} &
-s_{23} c_{12} -c_{23} s_{12} s_{13} e^{i\delta} &
c_{23} c_{13} 
\end{array} \right) ,
\eqno(2.16)
$$
where 
$(s_{12}, c_{12}) = (0.235, 0.974)$, 
$(s_{23}, c_{23}) = (0.0412, 0.999)$, 
$(s_{13}, c_{13}) = (0.00351, 1.000)$ and 
$\delta = + 72.2^\circ$~\cite{PDG14}.

For comparison with Case I, we consider an opposite extreme 
case (Case II): 
$$
 U^u  \simeq V_\text{CKM}^\dagger, \ \ \ \  U^d \simeq {\bf 1}, 
\ \ \ \ \ \ \ \ ({\rm Case\ I\hspace{-1pt}I}) , 
\eqno(2.17)
$$
although such case is not likely in the realistic quark mass matrix
model.

In addition to these cases, we investigate Case I\hspace{-1pt}I\hspace{-1pt}I, 
in which up- and down-quark mixings are sizable: 
$$
\begin{array}{l}
\tilde{U}^u = \left(
\begin{array}{ccc}  
0.999  
& 0.0320\, e^{i\, 8.14^\circ}  
& 0.0167\, e^{i\, 176^\circ} 
\\ 
0.0351\, e^{i\,172^\circ}  
& 0.970 
& 0.242\, e^{i\,168^\circ} 
\\ 
0.00845\, e^{i\,3.95^\circ} 
& 0.243\, e^{i\, 12.1^\circ}  
& 0.970
\end{array} \right) , 
\\
\tilde{U}^d = \left(
\begin{array}{ccc}
0.977 
& 0.212\,e^{i\,119^\circ}  
& 0.0126\,e^{i\,166^\circ} 
\\ 
0.207\,e^{i\,61.3^\circ} 
& 0.957 
& 0.203\,e^{i\,168^\circ} 
\\ 
0.0506\,e^{i\,60.8^\circ} 
& 0.197\,e^{i\,12.6^\circ} 
& 
0.979 
\end{array} \right) .
\end{array}
\eqno(2.18)
$$
The mixings in (2.18) have been derived in a mass matrix 
model~\cite{K-N_PRD15} which is notable one: 
a unified description of the quark- and lepton-mixing matrices 
and mass ratios has been described by using only the observed 
charged lepton masses as family-number dependent parameters.

It is worth investigating the potential of the $\mu$-$e$ 
conversion to determine the quark mixing. 
To do this, we consider Case I\hspace{-1pt}V with following 
parametrization: 
$$
U^u = R_3 \equiv \left( 
\begin{array}{ccc}
\cos\theta 
& \sin\theta 
& 0 
\\
-\sin\theta 
& \cos\theta 
& 0 
\\
0 
& 0 
& 1
\end{array} \right) , \ \ 
U_d = R_3^T V_{CKM} .
\eqno(2.19)
$$

\section{$\mu$-$e$ conversion in the FGB model}

We formulate the reaction rate of $\mu$-$e$ conversion in muonic 
atoms via $A_{2}^{\ 1}$ exchange based on Ref.~\cite{Kitano:2002mt}. 
Note that in the FGB model other muon lepton family violating (LFV) 
reactions   
($\mu \to e \gamma$, 
$\mu \to 3e$, 
$\mu^{-}e^{-} \to e^{-}e^{-}$ in muonic atom~\cite{Koike:2010xr}, 
and so on) arise at higher order. These reaction rates are 
suppressed by higher order couplings, gauge invariance, 
and so on. Hence we do not study these reactions here.

The $\mu$-$e$ conversion via $A_{2}^{\ 1}$ 
exchange is described by the effective interaction 
Lagrangian\footnote{
We omit the contribution via the kinetic mixing of 
$A_{2}^{\,1}$ and $Z$ boson. The contribution is suppressed 
by the loop factor and quark mixings, and is sub-dominant 
relative to direct ones of $A_{2}^{\ 1}$.}, 
$$
\begin{array}{ll}
\displaystyle{
	\mathcal{L}_\text{int} 
= 
	\left( \frac{g_{F}^{X}}{\sqrt{2}} \right)^{2}
	\frac{1}{M_{12}^{2}}  
	\sum_{q=u,d}
	\Bigl\{ 
	C_{L(q)}^{X,\alpha} 
	\left( \bar{e}_L \gamma^{\mu} {\mu}_L \right) 
	\left( \bar{q} \gamma_{\mu} {q} \right) 
	}
\displaystyle{
	C_{R(q)}^{X,\alpha} 
	\left( \bar{e}_R \gamma^{\mu} {\mu}_R \right) 
	\left( \bar{q} \gamma_{\mu} {q} \right)
	\Bigr\} 
}. 
\end{array}
	\label{Eq:Leff} 
\eqno(3.1)      
$$
Here $X$ and $\alpha$ denote the model, $X \ni 
\{\text{A}, \text{B}_{1}, \text{B}_{2} \}$, and the type of 
quark mixing matrices, 
$\alpha \ni \{\text{I}, \text{I\hspace{-1pt}I}, 
\text{I\hspace{-1pt}I\hspace{-1pt}I}, \text{I\hspace{-1pt}V} \}$, 
respectively. 
The coefficients $C_{L(q)}^{X,\alpha}$ and $C_{R(q)}^{X,\alpha}$ 
are derived from interaction Lagrangian in each model discussed in 
previous section. We list $C_{L(q)}^{X,\alpha}$ in the generation 
basis in Table 2. $C_{R(q)}^{X,\alpha}$ is related with 
$C_{L(q)}^{X,\alpha}$ as follows, 
$$
	C_{R(u)}^{X,\alpha}
	= 	\left\{
	\begin{array}{l}
	+ C_{L(u)}^{X,\alpha} ~~ \text{for } X=\text{A}, \\[2mm]
	- C_{L(u)}^{X,\alpha} ~~ \text{for } X=\text{B}_{1} 
	\text{ and } \text{B}_{2}, 
	\end{array}
	\right.
\eqno(3.2)
$$
$$
	C_{R(d)}^{X,\alpha} = 
	\left\{
	\begin{array}{l}
	+ C_{L(d)}^{X,\alpha} ~~ \text{for } X=\text{A}, \\[2mm]
	- C_{L(d)}^{X,\alpha} ~~ \text{for } X=\text{B}_{1} 
	\text{ and } \text{B}_{2}. 
	\end{array}
	\right.
\eqno(3.3)
$$

\begin{table}[t!] 
\caption{$C_{L(u)}^{X,\alpha}$ and $C_{L(d)}^{X,\alpha}$ 
for each model and for each quark mixing matrix. 
$V_{qq'}$ and $\tilde{U}^{q}$ stand for the CKM matrix and 
the mixing matrices derived in Ref.~\cite{K-N_PRD15}, 
respectively.}
\vspace{0mm}
\begin{center}
\begin{tabular}{llllllll}
\hline

& 
& Model A 
& Model $\text{B}_1$
& Model $\text{B}_2$ 
\\ \hline
& $C_{L(u)}^{X,\text{I}}$ \hspace{3mm}
& $0$ 
& $0$
& $0$
\\[0mm]

& $C_{L(d)}^{X,\text{I}}$
& $-V_{cd}^{*} V_{ud}$
& $-V_{cd}^{*} V_{td}$
& $-V_{ud}^{*} V_{td}$
\\[2mm]
& $C_{L(u)}^{X,\text{I\hspace{-1pt}I}}$
& $-V_{us} V_{ud}^{*}$
& $-V_{us} V_{ub}^{*}$ 
& $-V_{ud} V_{ub}^{*}$  
\\

& $C_{L(d)}^{X,\text{I\hspace{-1pt}I}}$
& $0$
& $0$
& $0$
\\[2mm]
& $C_{L(u)}^{X,\text{I\hspace{-1pt}I\hspace{-1pt}I}}$
& $-(\tilde{U}_{cu}^{u})^{*} \tilde{U}_{uu}^{u}$ 
& $-(\tilde{U}_{cu}^{u})^{*} \tilde{U}_{tu}^{u}$ 
& $-(\tilde{U}_{uu}^{u})^{*} \tilde{U}_{tu}^{u}$  
\\

& $C_{L(d)}^{X,\text{I\hspace{-1pt}I\hspace{-1pt}I}}$
& $-(\tilde{U}_{sd}^{d})^{*} \tilde{U}_{dd}^{d}$ 
& $-(\tilde{U}_{sd}^{d})^{*} \tilde{U}_{bd}^{d}$ 
& $-(\tilde{U}_{dd}^{d})^{*} \tilde{U}_{bd}^{d}$  
\\
\hline
\end{tabular} 
\label{table:couplings}
\end{center}
\end{table}

The branching ratio of $\mu$-$e$ conversion is defined by 
$\text{B} (\mu^{-} N \to e^{-} N) = 
\omega_\text{conv}/\omega_\text{capt}$, 
where $\omega_\text{conv}$ and $\omega_\text{capt}$ represent 
the reaction rates of $\mu$-$e$ conversion and of the muon capture 
process, respectively. 
The reaction rate $\omega_\text{conv}$ is calculated by the overlap 
integral of wave functions of the initial muon, the final electron, and the 
initial and final nucleus. In the FGB model, $\omega_\text{conv}$ 
is 
$$
\begin{array}{ll}
\displaystyle{
	\omega_\text{conv} 
	= 
	\left( \frac{g_{F}^{X}}{\sqrt{2}} \right)^{4}
	\frac{4 m_{\mu}^{5}}{M_{12}^{4}}
	\Bigl| 
	\left( 2C_{L(u)}^{X,\alpha} + C_{L(d)}^{X,\alpha} \right) V^{(p)} 
	} 
\displaystyle{
	+ \left( C_{L(u)}^{X,\alpha} + 2 C_{L(d)}^{X,\alpha} \right) V^{(n)} 
	\Bigr|^2 
	+ 
	\left( L \leftrightarrow R \right)
	}. 
\end{array}
\eqno(3.4)  	
$$
Here $m_{\mu}$ is the muon mass. 
The overlap integral of wave 
functions of muon, electron, and protons (neutrons) gives $V^{(p)}$ 
($V^{(n)}$) (explicit formulae and details of the calculation are 
explained in Ref.~\cite{Kitano:2002mt}).  
We list $V^{(p)}$ and $V^{(n)}$ for relevant nuclei of 
SINDRUM-I\hspace{-1pt}I (Au), DeeMe (C and Si), COMET (Al and Ti), 
Mu2e (Al and Ti), and PRISM (Al and Ti) in Table 3. 
We also list them for U nucleus. The $\mu$-$e$ conversion 
search with the U target can assist to confirm the FGB model and to 
determine the quark mixings.

\begin{table}[t!] 
\caption{
The overlap factor of wave functions and the muon 
capture rate $\omega_{capt}$ for each nucleus $N$. 
}
\vspace{0mm}
\begin{center}
\begin{tabular}{llllllll}
\hline
$N$ 
& $V^{(p)}$ ~~~~
& $V^{(n)}$ ~~~~
& $\omega_{capt}(s^{-1})$
\\ \hline 
C 
& $3.12 \times 10^{-3}$
& $3.12 \times 10^{-3}$
& $3.88 \times 10^{4}$
\\
Si
& $1.87 \times 10^{-2}$
& $1.87 \times 10^{-2}$
& $8.71 \times 10^{5}$
\\
Al
& $1.61 \times 10^{-2}$
& $1.73 \times 10^{-2}$
& $7.05 \times 10^{5}$
\\
Ti
& $3.96 \times 10^{-2}$
& $4.68 \times 10^{-2}$
& $ 2.59 \times 10^{6}$
\\
Au
& $9.74 \times 10^{-2}$
& $1.46 \times 10^{-1}$
& $ 1.31\times 10^{7}$
\\
U
& $7.98 \times 10^{-2}$
& $1.27 \times 10^{-1}$
& $1.24 \times 10^{7}$
\\ \hline
\label{Tab:V_caprate}
\end{tabular} 
\end{center}
\end{table}

\section{Numerical result}

\begin{table}[t!] 
\begin{center}
\caption{Lower bound on $M_{12}$ for each quark mixing in each 
model from the $\mu$-$e$ conversion limit at SINDRUM-I\hspace{-1pt}I, 
$\text{B} (\mu^{-} \text{Au} \to e^{-} \text{Au}) < 7 \times 
10^{-13}$~\cite{SINDRUM_EPJC06}. 
}
\vspace{1mm}
\begin{tabular}{llllllll}
\hline
Case \hspace{5mm}
& Model A \hspace{5mm}
& Model B$_1$ \hspace{5mm}
& Model B$_2$ \hspace{5mm}
\\ \hline 
I 
& 291\,TeV
& 24.2\,TeV 
& 50.4\,TeV
\\
I\hspace{-1pt}I 
& 273\,TeV 
& 14.4\,TeV 
& 29.8\,TeV 
\\
I\hspace{-1pt}I\hspace{-1pt}I
& 273\,TeV 
& 54.8\,TeV
& 126\,TeV
\\ \hline
\end{tabular} 
\end{center}
\end{table}

\begin{table}[t]
\begin{center}
\caption{
$\text{B}(\mu^- \text{Al} \to e^- \text{Al})$ for each Case and Model. 
The values are given in a unit of 
$n^{-2} (M_{12} / 10^3 {\rm TeV})^{-4}$ for Model A, and
$n^{-2} (M_{12} / 10^2 {\rm TeV})^{-4}$ for Model B. }
\vspace{1mm}
\begin{tabular}{lllll} \hline
Case \hspace{5mm}
&  Model A \hspace{5mm}     
&  Model B$_1$ \hspace{5mm}    
&  Model B$_2$ \hspace{5mm}
\\ \hline
I   
& $8.54 \times 10^{-17}$ 
& $7.54 \times 10^{-16}$ 
& $1.42\times 10^{-14}$ 
\\
I\hspace{-1pt}I  
& $1.51 \times 10^{-15}$ 
& $1.14 \times 10^{-16}$ 
& $2.22 \times 10^{-15}$ 
\\
I\hspace{-1pt}I\hspace{-1pt}I  
& $1.22 \times 10^{-15}$ 
& $1.94 \times 10^{-14}$ 
& $5.64 \times 10^{-13}$ 
\\ \hline 
\end{tabular}
\end{center}
\end{table}

\begin{figure}[t!]
\begin{center}
\includegraphics[clip, width=80mm]{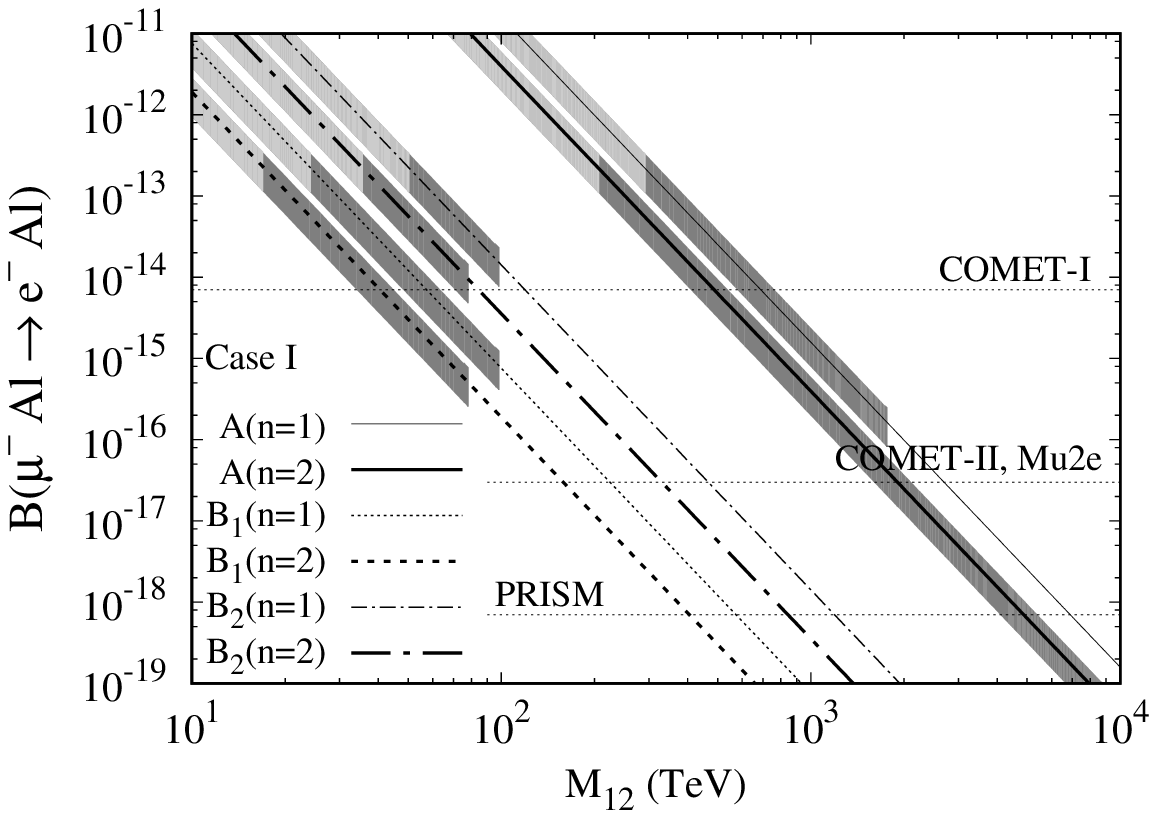}
\includegraphics[clip, width=80mm]{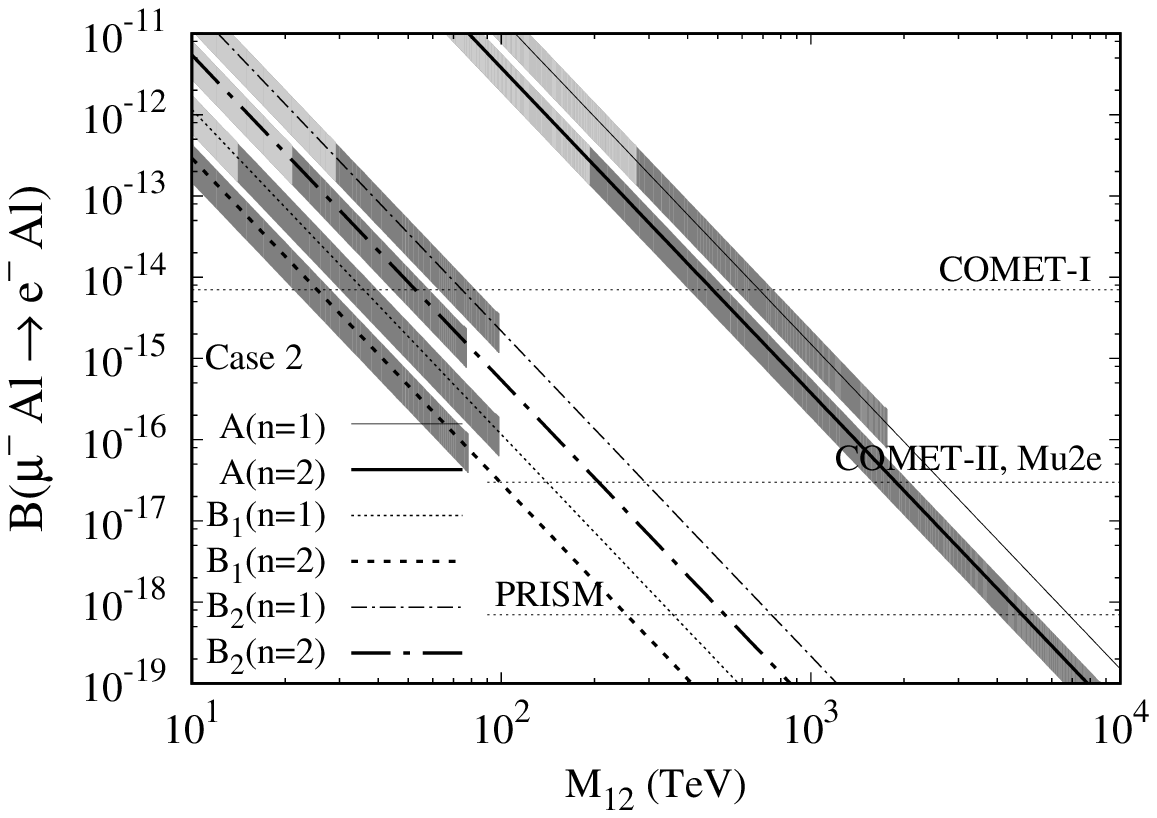}
\includegraphics[clip, width=80mm]{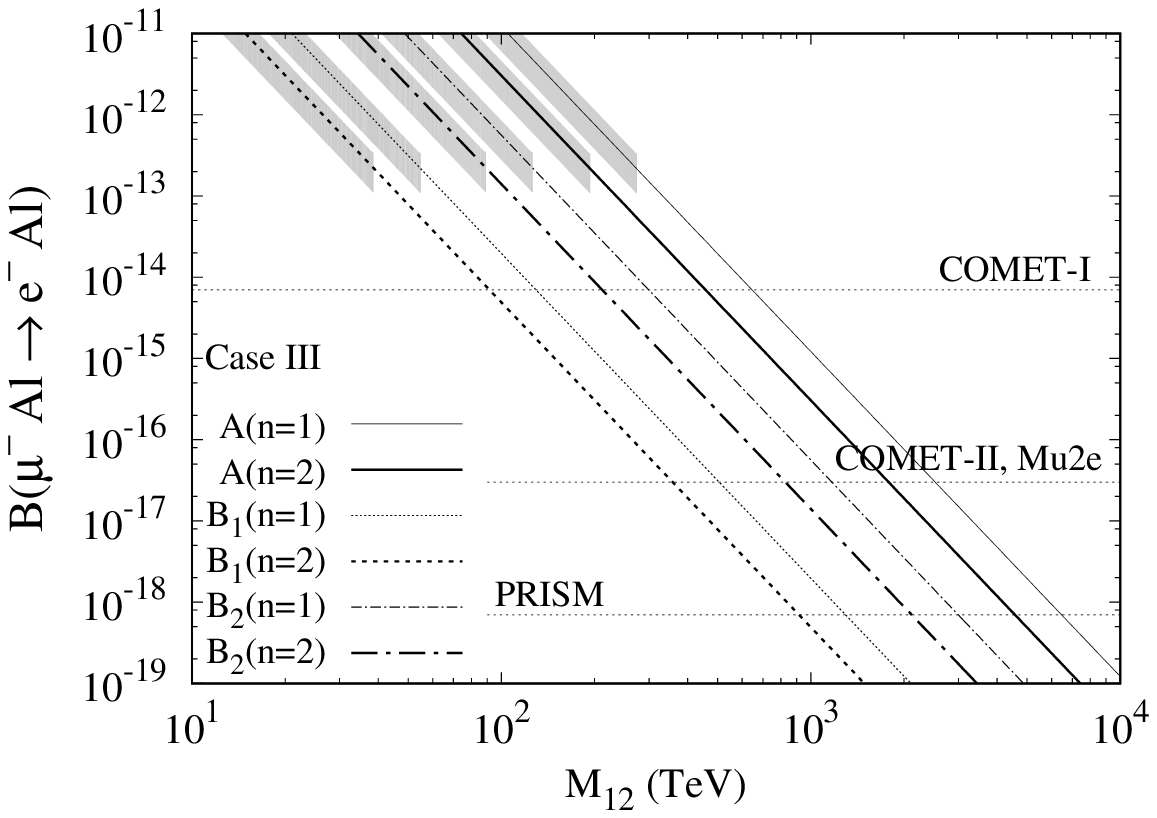}
\end{center}
\caption{$M_{12}$ dependence of $\text{B}(\mu^{-}\text{Al} 
\to e^{-}\text{Al})$ for $U_{u} = \boldsymbol{1}$ 
and $U_{d} = V_\text{CKM}$ (upper panel), for 
$U_{u} = V_\text{CKM}^{\dagger}$ and $U_{d} = 
\boldsymbol{1}$ (middle panel), and for 
$U^{u} = \tilde{U}^{u}$ and $U^{d} = \tilde{U}^{d}$ (see 
Eq.~(2.18)) (lower panel). 
Light and dark shaded region is excluded region by the 
SINDRUM-I\hspace{-1pt}I and by the observed 
$P^{0}$-$\bar{P}^{0}$ mixing (Table 1), respectively. 
Horizontal dashed lines show the single event sensitivities 
of each experiment (see Introduction). }
\label{Fig:M12dep_Al}
\end{figure}

\begin{table}[t]
\caption{
$\text{B}(N)/\text{B}(\text{Al})$ 
in each model and for each quark mixing matrix. 
In case I and I\hspace{-1pt}I, $\text{B}(N)/\text{B} 
(\text{Al})$ is universal for each model.
}
\vspace{-2mm}
\begin{center}
\begin{tabular}{llllllll}
\hline
$N$ 
& case I 
& case I\hspace{-1pt}I 
& case I\hspace{-1pt}I\hspace{-1pt}I (A)
& case I\hspace{-1pt}I\hspace{-1pt}I (B$_{1}$)
& case I\hspace{-1pt}I\hspace{-1pt}I (B$_{2}$)
\\ \hline
Ti 
& 1.88
& 1.77 
& 1.88
& 1.88
& 1.87
\\[0mm]
C
& 0.620
& 0.650
& 0.619
& 0.619
& 0.623
\\[0mm]
Si
& 0.991
& 1.040
& 0.990 
& 0.990
& 0.996  
\\
Au 
& 3.18
& 2.56
& 3.21
& 3.20
& 3.12
\\[0mm]
U 
& 2.47
& 1.91 
& 2.49 
& 2.48
& 2.41  
\\
\hline
\end{tabular} 
\label{table:ratio}
\end{center}
\end{table}

We are now in a position to show numerical results. 
Table 4 shows the lower bound on the FGB mass $M_{12}$ by 
the $\mu$-$e$ conversion search at SINDRUM-I\hspace{-1pt}I, 
$\text{B} (\mu^{-} \text{Au} \to e^{-} \text{Au}) < 7 
\times 10^{-13}$~\cite{SINDRUM_EPJC06}. 
Current most stringent limits of $M_{12}$ are obtained from 
observed $P^{0}$-$\bar{P}^{0}$ oscillations (Table 1), not 
from the $\mu$-$e$ conversion search.

Next we show the feasibility of FGB search in $\mu$-$e$ 
conversion search experiments. Fig.~\ref{Fig:M12dep_Al} 
shows $\text{B} (\mu^{-} \text{Al} \to e^{-} \text{Al})$ 
as a function of $M_{12}$. 
In light of the cancellation, the FGB masses are supposed to be 
up to $\sim 10^{4}\,\text{TeV}$~\cite{Sumino_PLB09, 
Sumino_JHEP09} (see Sec.~2). 
As is shown in Fig.~\ref{Fig:M12dep_Al}, next generation 
experiments cover most of this mass region, and the discovery 
of $\mu$-$e$ conversion via $A_2^{\ 1}$ exchange is 
expected in near future. To put it the other way around null results of 
$\mu$-$e$ conversion search can rule out the FGB model.

After the discovery of $\mu$-$e$ conversion, we need to 
check whether the observed event is a signal of $A_2^{\ 1}$ or not. 
Table 6 lists the ratio of branching ratios, 
$\text{B}(\mu^{-} N \to e^{-} N)/\text{B}(\mu^{-} 
\text{Al} \to e^{-} \text{Al})$. 
The $\mu$-$e$ conversion events will be confirmed as the 
signal of $A_{2}^{\ 1}$ through precise measurements of 
the ratios. Also, a type of quark mixing matrix can be identified 
by the precise measurements. 
The $\mu$-$e$ conversion search by using large nucleus 
target is important. Indeed, although it is hard to 
distinguish the case I and I\hspace{-1pt}I\hspace{-1pt}I($A$) 
from the ratios 
$\text{B}(\text{Ti})/\text{B}(\text{Al})$, 
$\text{B}(\text{C})/\text{B}(\text{Al})$, 
and $\text{B}(\text{Si})/\text{B}(\text{Al})$, 
it can be possible for large nucleus, i.e., 
$\text{B}(\text{Au})/\text{B}(\text{Al})$, 
and $\text{B}(\text{U})/\text{B}(\text{Al})$. 
It is probably impossible to distinguish the case 
I\hspace{-1pt}I\hspace{-1pt}I($A$) and 
I\hspace{-1pt}I\hspace{-1pt}I($B_{1}$) from the ratios. 
To do this, we need additional observables via the FGB exchange, 
e.g., LFV kaon decays, LFV collider signals, and so on. 
Some of experiments are running or will launch in near future to 
search for these signals~\cite{Sher:2005sp, mueLHC}. Therefore 
it is important to simulate what correlations are expected and how 
sensitivity is required for the purpose. It is however beyond the 
scope of this paper and we leave them in future work~\cite{future}.

One may wonder why, in Table 6, $\text{B}(N)/\text{B} 
(\text{Al})$ is insensitive to Model in Case I and I\hspace{-1pt}I. 
This is understood as follows. 
The branching ratios can be decomposed into Model independent 
and dependent part as 
$$
	\text{B} 
	\propto 
	\frac{1}{n^{2}} 
	\frac{\bigl( g_{F}^{X} \bigr)^{4}}{M_{12}^{4}}
	\left| C_{L (d)}^{X, \text{I}} \right|^{2}
	\left| V^{(p)} + 2V^{(n)} \right|^{2} 
	\hspace{10mm} (\text{Case I}), 
	\eqno(4.1)	
$$
$$
	\text{B} 
	\propto 
	\frac{1}{n^{2}} 
	\frac{\bigl( g_{F}^{X} \bigr)^{4}}{M_{12}^{4}}
	\left| C_{L (u)}^{X, \text{I\hspace{-1pt}I}} \right|^{2}
	\left| 2V^{(p)} + V^{(n)} \right|^{2} 
	\hspace{10mm}  (\text{Case I\hspace{-1pt}I}). 
	\eqno(4.2)	
$$
For any target nuclei, the Model dependent part 
$\bigl( g_{F}^{X} \bigr)^{4} \bigl| 
C_{L (d)}^{X, \text{I}} \bigr|^{2}$ and 
$\bigl( g_{F}^{X} \bigr)^{4} \bigl| 
C_{L (u)}^{X, \text{I\hspace{-1pt}I}} \bigr|^{2}$ are  
cancelled in the ratio $\text{B}(N)/\text{B} (\text{Al})$.
Hence, in Case I and I\hspace{-1pt}I, the change in model 
does not affect the ratio.

\begin{figure}[t!]
\begin{center}
\includegraphics[clip, width=80mm]{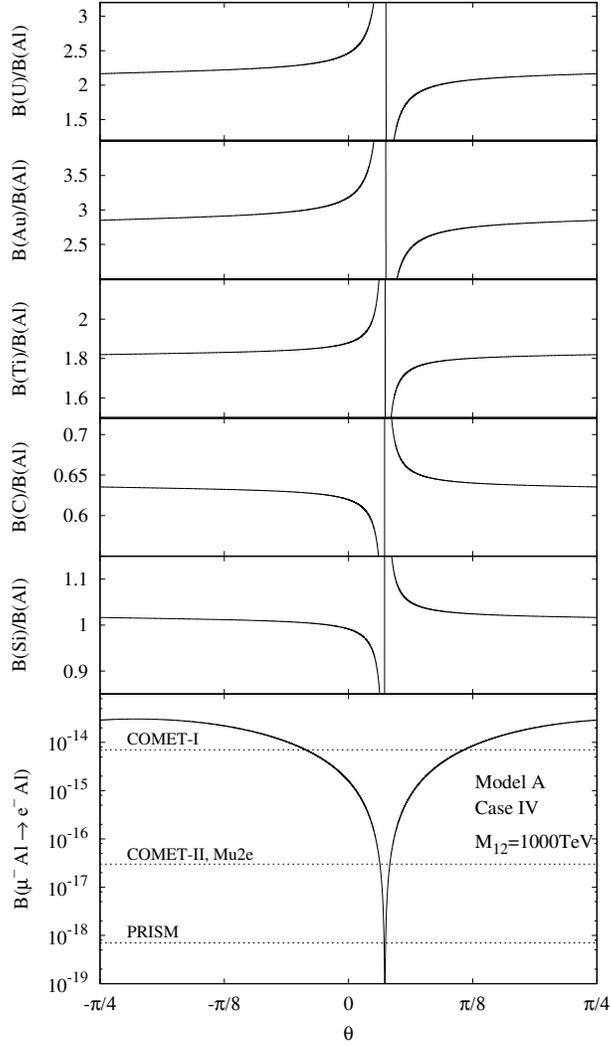}
\end{center}
\caption{$\theta$ dependence of $\text{B}(\mu^{-} \text{Al} \to 
e^{-} \text{Al})$ and of ratios $\text{B}(\mu^{-} \text{N} \to 
e^{-} \text{N})/\text{B}(\mu^{-} \text{Al} \to e^{-} \text{Al})$ 
in the model A. We took $M_{12} = 1000\,\text{TeV}$. 
Horizontal dashed lines show the single event sensitivities of each 
experiments (see Introduction)}
\label{Fig:theta_dep_modA_n1_1000TeV}
\end{figure}

\begin{figure}[t!]
\begin{center}
\includegraphics[clip, width=80mm]{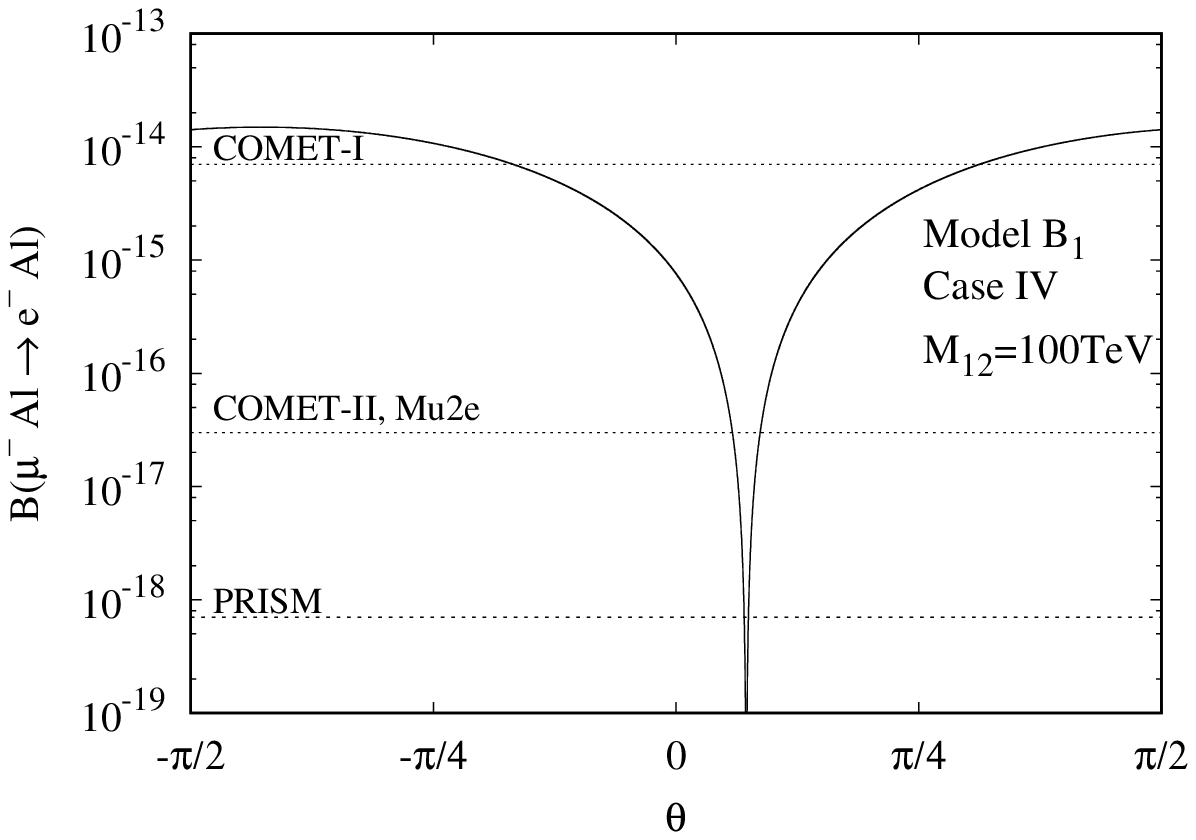}
\includegraphics[clip, width=80mm]{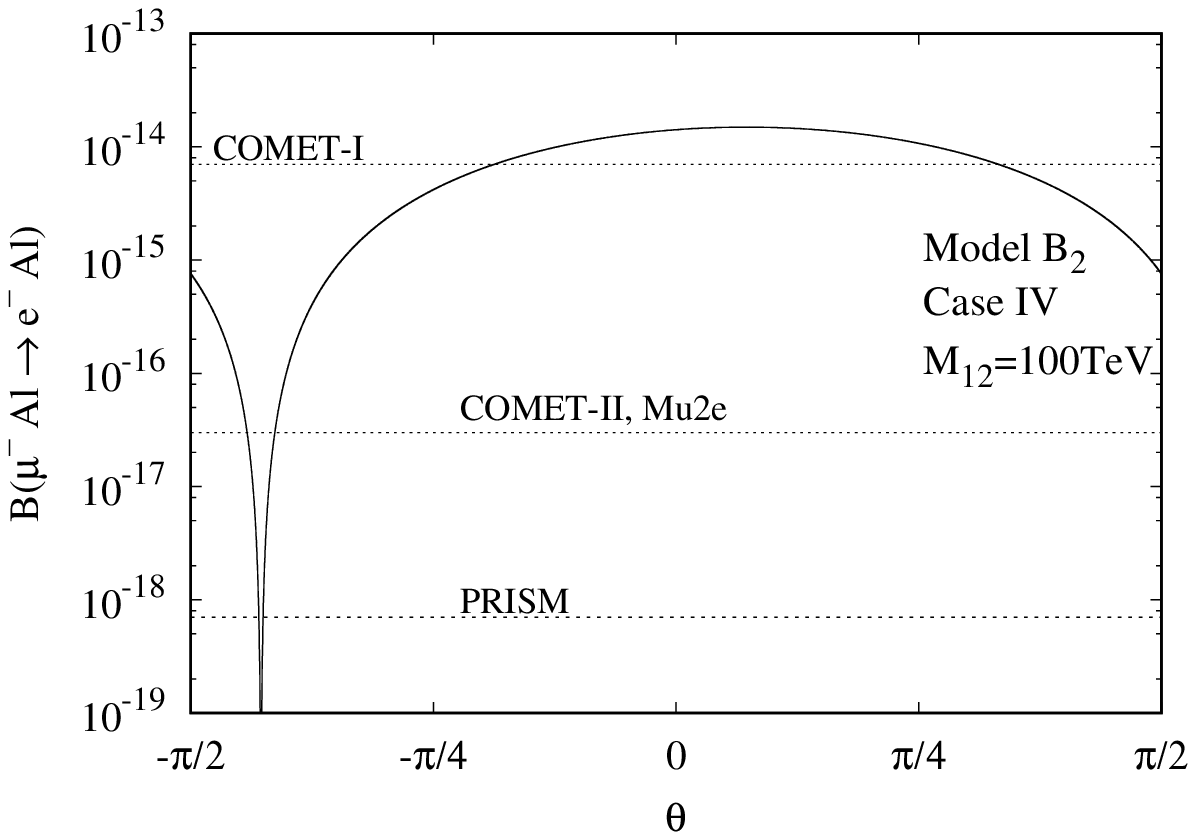}
\end{center}
\caption{$\theta$ dependence of $\text{B}(\mu^{-} \text{Al} 
\to e^{-} \text{Al})$ in the model $\text{B}_{1}$ (upper plot) 
and in the model $\text{B}_{2}$ (lower plot). 
Horizontal dashed lines show the single event sensitivities of each 
experiments (see Introduction). }
\label{Fig:theta_dep_modB_n1_100TeV}
\end{figure}

Finally we discuss the determination of the quark mixing  
by using parametrized mixing matrix~(2.19). 
The $\theta$ dependence of $\text{B}(\mu^{-} \text{Al} \to 
e^{-} \text{Al})$ is plotted in 
Figs.~\ref{Fig:theta_dep_modA_n1_1000TeV} 
(Model A, $M_{12} = 1000\,\text{TeV}$) and 
\ref{Fig:theta_dep_modB_n1_100TeV} (Model B$_{1}$ and  
B$_{2}$, $M_{12} = 100\,\text{TeV}$), respectively. 
The ratios $\text{B} (N)/\text{B} (\text{Al})$ as a function 
of $\theta$ are also shown in 
Fig.~\ref{Fig:theta_dep_modA_n1_1000TeV}. 
The results at $\theta = 0$ corresponds to those in Case I. 
The structure of mixing matrix can be determined through the 
precise measurement of $\text{B}(\mu^{-} \text{Al} \to 
e^{-} \text{Al})$. 
Particularly, in Model A, since the ratios $\text{B} (N)/\text{B} 
(\text{Al})$ also depends on $\theta$, the quark mixing can be 
accurately determined by accumulating a large number of 
$\mu$-$e$ conversion events.
Fig.~\ref{Fig:theta_dep_modA_n1_1000TeV} emphasizes an 
importance of the $\mu$-$e$ conversion searches with various 
target nuclei. In Model A, even if the signal of $\mu^{-} \text{Al} 
\to e^{-} \text{Al}$ will never be found, a number of events can 
be observed at experiments with other target nucleus. 
On the other hand, in Models B$_{1}$ and B$_{2}$, the ratios 
$\text{B} (N)/\text{B} (\text{Al})$ are independent of $\theta$, 
and are equal to those of Case I. This is because that the branching 
ratios can be decomposed into $\theta$ dependent part and independent 
part as follows 
$$
	\text{B}_{(\text{B}_{1})} 
	\propto 
	\bigl| V^{(p)} + 2V^{(n)} \bigr|^{2} 
	\bigl| 
	\bigl( V_{ud}^{*} \sin\theta + 
	V_{cd}^{*} \cos\theta \bigr) 
	V_{td} 
	\bigr|^{2},  
	\eqno(4.3)
$$
$$
	\text{B}_{(\text{B}_{2})} 
	\propto
	\bigl| V^{(p)} + 2V^{(n)} \bigr|^{2} 
	\bigl| 
	\bigl( V_{ud}^{*} \cos\theta - 
	V_{cd}^{*} \sin\theta \bigr) 
	V_{td} 
	\bigr|^{2} , 
\eqno(4.4)
$$
and the $\theta$ dependent part is canceled in $\text{B} 
(N)/\text{B} (\text{Al})$. 
Thus, in Models $\text{B}_{1}$ and $\text{B}_{2}$, it is 
difficult to examine the structure by the $\mu$-$e$ conversion search 
only. In such a case, it is necessary to combine the $\mu$-$e$ 
conversion search with other observables.

\section{Concluding remarks}

We have investigated the $\mu$-$e$ conversion  
via an exchange of family gauge boson $A_2^{\ 1}$ in a 
$U(3)$ FGB model. 
In the model there are various types of FGB spectrum and of 
family-number assignments. We have considered three 
well-motivated models: a model with inverted family-number 
assignment (Model A), and models with 
twisted ones (Model B$_{1}$ and B$_{2}$).   
We also have a degree of freedom of choice of quark mixing 
$U^u$ and $U^d$. We have introduced four types of mixing:  
a most likely mixing, $U^u  \simeq {\bf 1}$ and $U^d \simeq 
V_\text{CKM}$ (Case I), 
an opposite type of Case I, $U^u  \simeq V_\text{CKM}^\dagger$ 
and $U^d \simeq {\bf 1}$ (Case I\hspace{-1pt}I), 
a phenomenologically derived mixing (2.18), 
$U^u \simeq \tilde{U}^{u}$ and $U^d \simeq \tilde{U}^{d}$ 
(Case I\hspace{-1pt}I\hspace{-1pt}I), 
and a parametrized mixing, $U^{u} = R_{3}$ and $U^{d} = 
R_{3}^{T} V_\text{CKM}$ (Case I\hspace{-1pt}V).

We have calculated the branching ratio of $\mu$-$e$ conversion 
process, $\text{B}(\mu^{-} N \to e^{-} N)$, in Models A, 
B$_{1}$ and B$_{2}$ for each type of quark mixing.  
We have shown that next generation $\mu$-$e$ conversion search 
experiments will cover entire energy scale of the FGB model, and 
could confirm or rule out the FGB model. 
Muon-number violating decays except for the $\mu$-$e$ conversion 
is extremely suppressed in the FGB model. 
Thus we have emphasized the importance of precise measurements 
of the ratios $\text{B} (N)/\text{B} (\text{Al})$, which is necessary 
to confirm the FGB model. Searches for LFV decays of mesons  
should 
assist the confirmation. This interesting possibility is left for future 
work~\cite{future}. 

In the FGB model it is, in principle, possible to individually 
determine quark mixing matrix $U^u$ and $U^d$, in contrast 
within the SM. 
However, since $V^{(p)} \simeq V^{(n)}$ in the most nuclei, 
it is hard to observe the difference between $U^u$ and $U^d$. 
We hope that further precise search for the $\mu$-$e$ conversion
with heavy nuclei, e.g., Au and/or U which $V^{(p)}$ and 
$V^{(n)}$ are sizably different.

\section*{Acknowledgments}   

This  work was supported by the Grants-in-Aid for Scientific Research 
(KAKENHI) Grant No.\,16K05325 (Y.K) and No.\,16K05325 and 
No.\,16K17693 (M.Y.).


\end{document}